\documentclass[aps,footinbib,twocolumn,showpacs,amsmath,amssymb,superscriptaddress,prb,groupedaddress]{revtex4}
\usepackage{graphicx,amsmath}
\usepackage{epstopdf}
\usepackage{amssymb}
\usepackage{grffile}
\usepackage[usenames]{color}
\usepackage{indentfirst}
\usepackage{float}
\usepackage{color}
\usepackage{mathrsfs}
\usepackage{dcolumn}
\usepackage{bm}
\usepackage[colorlinks=true,bookmarks=false,citecolor=blue,linkcolor=red,urlcolor=blue]{hyperref}

\begin{document}
\title{ Fractional quantum Hall  interface induced by geometric singularity }

\author{Qi Li}
\email{liqi@aircas.ac.cn}
\affiliation{GBA Branch of Aerospace Information Research Institute, Chinese Academy of Sciences, Guangzhou 510700, People's Republic of China}
\affiliation{Guangdong Provincial Key Laboratory of Terahertz Quantum Electromagnetics, Guangzhou 510700, People's Republic of China}

\author{Yi Yang$^{a}$}
\thanks{$^{a}$These authors have contributed equally to this work.}
\affiliation{Department of Physics and Chongqing Key Laboratory for Strongly Coupled Physics, Chongqing University, Chongqing 401331, People's Republic of China}

\author{Zhou Li}
\affiliation{GBA Branch of Aerospace Information Research Institute, Chinese Academy of Sciences, Guangzhou 510700, People's Republic of China}
\affiliation{Guangdong Provincial Key Laboratory of Terahertz Quantum Electromagnetics, Guangzhou 510700, People's Republic of China}
\affiliation{University of Chinese Academy of Sciences, Beijing 100039, People's Republic of China}

\author{Hao Wang}
\affiliation{Shenzhen Institute for Quantum Science and Engineering, Southern University of Science and Technology, Shenzhen 518055, People's Republic of China}

\author{Zi-Xiang Hu}
\email{zxhu@cqu.edu.cn}
\affiliation{Department of Physics and Chongqing Key Laboratory for Strongly Coupled Physics, Chongqing University, Chongqing 401331, People's Republic of China}

\date{\today}
\begin{abstract}
The geometric response of quantum Hall liquids is an important aspect to understand their topological characteristics in addition to the electromagnetic response. According to the Wen-Zee theory, the topological spin is coupled to the curvature of the space in which the electrons reside. The presence of conical geometry provides a local isolated geometric singularity, making it suitable for exploring the geometric response. In the context of two-dimensional electrons in a perpendicular magnetic field, each Landau orbit occupies the same area. The cone geometry naturally provides a structure in which the distances between two adjacent orbits gradually change and can be easily adjusted by altering the tip angle. The presence of a cone tip introduces a geometric singularity that affects the electron density and interacts with the motion of electrons, which has been extensively studied. Furthermore, this type of geometry can automatically create a smooth interface or crossover between the crystalline charge-density-wave state and the liquid-like fractional quantum Hall state. In this work, the properties of this interface are studied from multiple perspectives, shedding light on the behavior of quantum Hall liquids in such geometric configurations.
\end{abstract}
\maketitle

\section{Introduction}
Fractional quantum Hall (FQH) effects have revealed a range of exotic topologically ordered phases since its discovery~\cite{Tsui} for more than three decades. As an emergent phenomenon arised from interacting two-dimensional electron system with perpendicular magnetic field, numerous theoretical and experimental investigations are devoted to it. The first seminal work is contributed by Laughlin who gave an elegant trial wave function describing partial filling $\nu = 1/3$ state in the lowest Landau level (LLL)~\cite{Laughlin} which was proved to have fractional excitation and statistics. More exotic FQH states such as Moore-Read-like state at  half filling in the first Landau level with $\nu = 5/2$ are found to host non-Abelian topological excitations and statistics~\cite{Willett, Moore}.  Besides the regular descriptions of a quantum Hall system from electro-magnetic response, the topological state also has response to the geometric manifold where the electrons lives in. Such as the FQH state on a torus has topological degenerate and that on a sphere has a topological shift. The geometric responses include the anomalous viscosity~\cite{Avron, Levay, Read} and the gravitational anomaly~\cite{bradlyn, Can, Abanov} are less well-know but are topological characteristics of the QH state.  Haldane pointed out that the internal geometrical degree of freedom to the change of the correlation hole shape is responsible for the dynamical variation of the guiding-center metric~\cite{Haldane, Haldane1}. Following earlier seminal work by Wen and Zee~\cite{Wen}, the response of FQH states to changes in spatial geometry and topology, such as points of singular curvature in real space or geometry with different genus, has been devoted to more efforts~\cite{Can1, Biswas, Ying}.

Recently, experimental efforts are devoted to creating synthetic materials in artificial magnetic fields such as for cold atoms and photons~\cite{Gemelke,  Aidelsburger, Kennedy, Miyake, Aidelsburger1, Doko, Lin, Juzeliunas, Peter}. Ultracold atomic gases in a fast rotating trap could be employed to the study of quantum Hall phases and transitions as one can precisely control the dipole-dipole interaction in an anisotropic way~\cite{Lu, Ni, Cooper, Fetter, Baranov, Osterloh, Baranov1, Qiu, Hu}. Likewise, artificial gauge fields could also be generated for photons. The Landau levels and even Laughlin type FQH state for photons is actualized~\cite{Mittal, Schine16, Otterbach, Juzeliunas1,Schine19,Clark20}. In experiment~\cite{Schine16, Schine19, Clark20}, photons are confined in a plane with several copies. Each copy is actually confined in a conical geometry. It not only provides the trap stability against centrifugal limit but also constructing point-like curved space with non-zero curvature at the tip.  The gravitational anomaly has already been extracted from the particle density near the cone tip as coupling to the local curvature~\cite{Can1, Ying}. Three topological quantities,   Chern number,  mean orbital spin and chiral central charge are measured through local electromagnetic and gravitational responses~\cite{Schine19}. Due to the holomophism feature of FQH wave function, the radial direction length of a cone manifold extends gradually accompanied with decreasing the cone angle, namely increasing the number of copies in experiment. The interval between two adjacent Landau orbits are thus increased and less overlapped. In this geometry, the change rate of  the  intervals  is inhomogeneous since each Landau orbit occupies a fixed area $2\pi\l_B^2$. Therefore, the Landau orbits near the tip  are far apart from each other compare to that near the edge. Similar to the Tao-Thouless(TT) state formed in the cylinder geometry, the electrons tend to form a crystalline TT state in thin cylinder limit. Because of the inhomogeneous change rate of the intervals of Landau orbits, the TT state is firstly formed near the cone tip and thus an smooth interface, or a crossover naturally emerges in the bulk separating the crystalline phase and FQH phase without artificial ``cut-and-glue"  operations ~\cite{Zhu} or designing a double quantum well systems ~\cite{Yang}. 

In this work, we investigate several properties of FQH states on a cone with the help of Jack polynomials and Monte Carlo simulation~\cite{Khanna}. 
The rest of this paper is arranged as follows. In Sec.\uppercase\expandafter{\romannumeral2} we briefly introduce the single particle eigenstates on cones. The ground state wave function of the many-body Hamiltonian of  FQH systems  could be obtained numerically using exact diagonalization (ED) or Jack polynomials method. Sec.\uppercase\expandafter{\romannumeral3}  gives the density profile and charge distribution for two typical FQH states on cones. The orbitial angular momentum calculations reveal the gravitational anomaly as response to curvature singularity  and the low-energy spectrum shows opposite chirality near the induced interface. We perform calculations with respect to wave function overlap and pair correlation functions based on conical wave function profiles in Sec.\uppercase\expandafter{\romannumeral4}.  In addition, we investigate the entanglement entropy in momentum space to record the formation of interface and manipulate the bipartition of the system in real space with an exact cutting position which could efficiently experience the singular curvature in real space.  Conclusion and discussion are presented in Sec.\uppercase\expandafter{\romannumeral5}. Some technical details of Monte Carlo simulation are given in the two appendices.

\section{Model and Method}

\begin{figure}[htbp]
\center
\includegraphics[width=8.5cm]{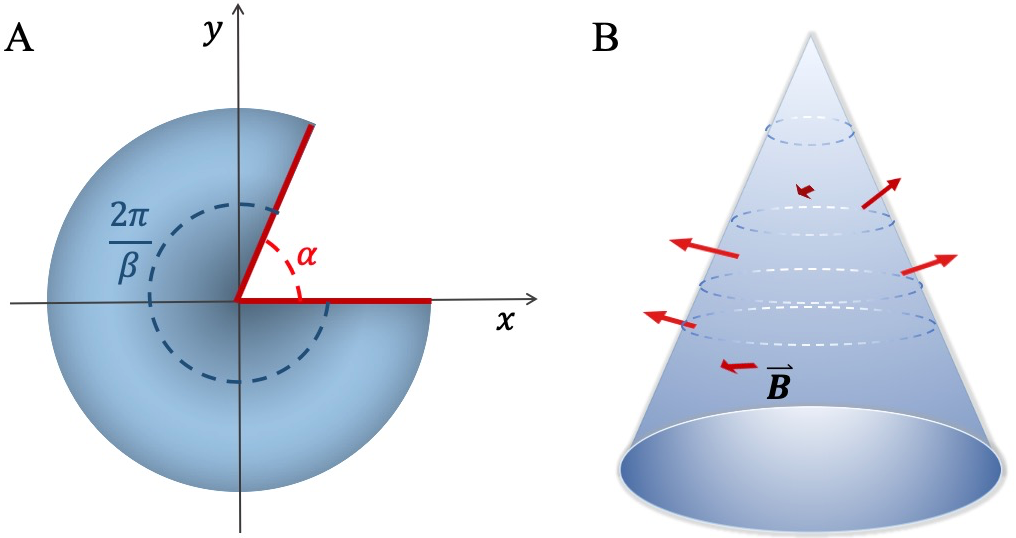}
\caption{The sketch of a cone built from a planar disk. (A)The cone mapped to a plane with deficit angle $\alpha$ and the remaining part with angle $\frac{2\pi}{\beta} = 2\pi - \alpha$.   (B) The three dimension perspective of a cone with a uniform magnetic field $\vec{B}$ penetrating the surface and single particle orbitals are formed with  symmetric gauge on a cone.}
\label{fig:sketch}
\end{figure}

As sketched in Fig.~\ref{fig:sketch},  the construction of isolated points with singular spatial curvature can be achieved by removing a sector with a specific apical angle from a disk and then gluing the resulting edges together. This creates a conical geometry with a point of singular curvature at the cone tip in real space, making it a relevant platform for physical studies. These points may be feasibly created within lattice systems experimentally. The curvature of a cone exhibits singularity only at the tip in the Gaussian curvature field but vanishes elsewhere ~\cite{Wen, Can, Biswas, Cho}. The Gauss-Bonnet theorem ensures that the integrated curvature enclosing the apex of the cone is related to the deficit angle $\alpha$. 
 
\begin{equation}
\int K(r) dS = \alpha = 2\pi(1-\frac{1}{\beta})
\end{equation}   

The two-dimensional charge carriers on the surface of a cone which is penetrated by uniform magnetic field $\vec{B}$  have effective single particle Hamiltonian:
\begin{equation}
H_0 = \frac{1}{2m}(\textbf{P} + e \textbf{A})^2
\end{equation}
Under symmetric gauge $\textbf{A} = (-By/2, Bx/2)$, one could write down the eigenstate wavefunction of Hamiltonian $H_0$ similar to the form in disk geometry. In general, there are two types of single particle wavefunctions ~\cite{Bueno} and the type-{\uppercase\expandafter{\romannumeral1}}  wavefunction can be written as follows:
\begin{equation}
\label{eq:wf}
\Phi_{n,m}^{\uppercase\expandafter{\romannumeral1}} (z) = \mathcal{N}_{n,m} z^{\beta m} L_n^{\beta m}(\vert z \vert^2 / 2) e^{-\vert z \vert ^2 /4}
\end{equation} 
where the complex coordinate $ z = (x+iy)/\ell_B = \vert z \vert e^{ i \theta}  $ with arg($z$) $ = \theta \in \left[ 0, 2\pi/ \beta  \right] $.  We set magnetic length $\ell_B = \sqrt{\hbar c / e B}  = 1 $. $L_n^{\beta m}(\vert z \vert^2 / 2)$  is generalized Laguerrel polynomial.  Here we could separate the angular and radial variations part of the eigenstate wavefunction into:
\begin{equation}
\Phi( \vert z \vert, \theta ) = \phi(\vert z \vert) e^{i\beta m \theta} 
\end{equation}
and the periodic boundary condition $\Phi(\vert z \vert, 2\pi/\beta) = \Phi(\vert z \vert, 0)$ comes from the gluing operation. The corresponding  type-{\uppercase\expandafter{\romannumeral1}}   eigenvalues 
\begin{equation}
E^{\uppercase\expandafter{\romannumeral1}}_{n, m} = (n + 1/2) \hbar \omega_c 
\end{equation}
with $n, m = 0, 1, 2, \cdots$ are independent of $m$ and responsible for the macroscopic degeneracy of the LLs.
The type-{\uppercase\expandafter{\romannumeral2}} eigenstates   
\begin{equation}
\Phi_{n,m}^{\uppercase\expandafter{\romannumeral2}} (z) = \mathcal{N}_{n,m} z^{\ast \beta m} L_n^{\beta m}(\vert z \vert^2 / 2) e^{-\vert z \vert ^2 /4}
\end{equation}
have eigenvalues 
\begin{equation}
E^{\uppercase\expandafter{\romannumeral2}}_{n, m} = (n + \beta m + 1/2) \hbar \omega_c 
\end{equation}
with $n = 0, 1, 2,\cdots$ and $m = 1, 2,\cdots$ which are related to $m$ and $\beta$. The normalization factor is:
\begin{equation}
\mathcal{N}^2_{n, m} = \frac{\beta n!} {  2\pi 2^{\beta m}  \Gamma (n + \beta m +1)}.
\end{equation}

When parameter $\beta = 1$, i.e., the flat disk case, LL index for type-{\uppercase\expandafter{\romannumeral1}} states is given by $n_{LL}  = n$ but $n_{LL} = n + m$ for type-{\uppercase\expandafter{\romannumeral2}} states  and both cases are degenerate. When parameter $\beta > 1$, i.e., a general cone case,  type-{\uppercase\expandafter{\romannumeral1}} states remain unchanged while energies of  type-{\uppercase\expandafter{\romannumeral2}} states elevate to the internal levels inside the inter-LL gaps. 
The states in the LLL come from type-{\uppercase\expandafter{\romannumeral1}}  with energy $E = \frac{1}{2}\hbar \omega_c$ and we will use the single particle  wavefunction which  refers to type-{\uppercase\expandafter{\romannumeral1}} $\Phi_{0LL} = \mathcal{N}_{0,m} z ^ {\beta m } \exp(-\vert z \vert^2/4)$   in subsequent parts   with no superscript any more. 

The Laughlin state at $\nu = 1/3$ can be obtained by diagonalizing the hard-core model Hamiltonian with $V_1$ Haldane's pseudopotential ~\cite{Laughlin, Haldane2}. It is known that the model wave functions could also be obtained with the help of Jack polynomial which is characterized by a root configuration and a parameter $\alpha$~\cite{Bernevig, Bernevig1, Bernevig2}. For example, the root configuration for Laughlin state is  ``$1001001\cdots$" and ``$11001100\cdots$"  for Moore-Read (MR) state~\cite{Moore}.  The leftmost orbit represents the innermost Landau orbit which could be the center  of a flat disk or the cone tip.  In addition to the ground state, one can similarly describe quasihole states with one addition unoccupied orbit at the cone tip, namely ``$01001001\cdots$".  In general, it is straightforward to consider Laughlin's model wave function $\Psi_L(z_1, \cdots, z_N) = \prod_{j<k}(z_j-z_k)^m e^{-\frac{1}{4}\sum_i \vert z_i \vert^2}$  for a state with a single quasihole located at $z_0$, 
$\Psi_{qh}(z_0) = \prod_i (z_i-z_0)\Psi_L(z_1, \cdots, z_N)$.  With model wave functions, it is also easy to simulate these FQH states by Metropolis Monte Carlo method.

\section{charge density profiles}
\begin{figure}[htbp]
\center
\includegraphics[width=8.5cm]{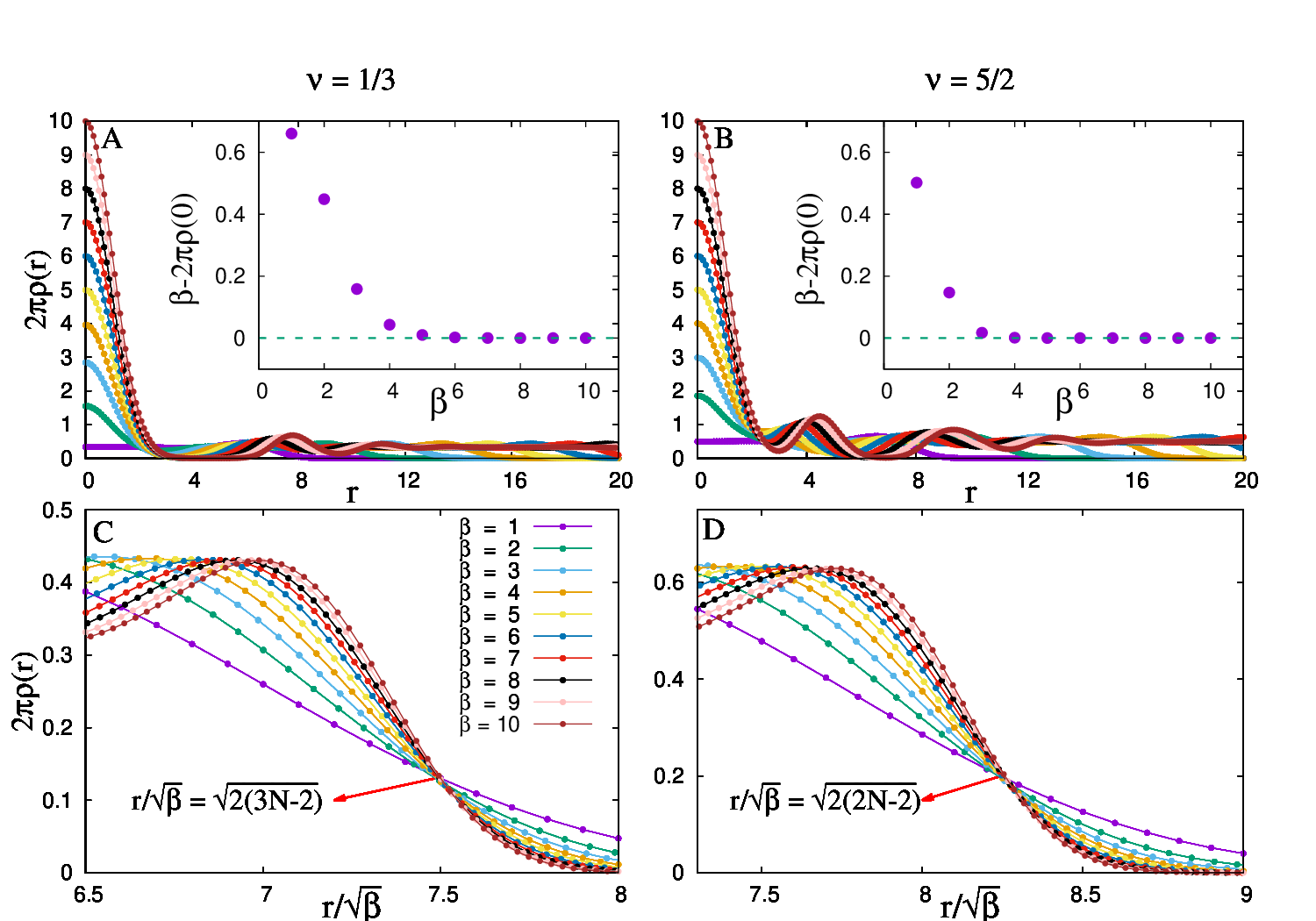}
\caption{The density profile $2\pi \rho(r)$ as a function of $r$ for $\nu = 1/3$ Laughlin state (A) and  $\nu = 5/2$ MR state (B) on cones. The inset plots in (A) and (B) show $\beta - 2\pi \rho(0)$ versus $\beta$ with respect to the corresponding states, respectively.   (C)(D) zoom in the density profile near the physical edge $r = \sqrt{2N_{orb}\beta}$ in rescaled coordinate axis $r/\sqrt{\beta}$.  The system size for Laughlin state (A)(C) is 10 electrons and for MR state (B)(D) is 18 electrons.}
\label{fig:density}
\end{figure}

The incompressible topological FQH ground state has a uniform bulk density $\rho_0=\nu / 2\pi \ell_B^2 $ at filling $\nu$ in smooth space such as infinite plane.  The density is nonuniform in the presence of a quantum Hall edge or interface.  Moreover, in a curved space,  the density has an extra correction as follows
\begin{equation}
\rho = \frac{\nu}{2\pi \ell_B^2} + \frac{\nu K(r)}{8\pi}(\mathcal{S} - 2j)
\end{equation}
with including the  Gaussian curvature $K(r)$,  the particle spin $j$~\cite{Can2}  and also the topological shift  $\mathcal{S}$~\cite{Wen}.  The topological shift $\mathcal{S} = \nu^{-1}$ for fermionic Laughlin state and $\mathcal{S} = \nu^{-1}+1$ for fermionic Moore-Read state.  In spherical geometry, the curvature is uniform, resulting in a constant correction everywhere, and thus the charge density remains uniform. However, in conical geometry, non-zero curvature emerges at the apex, leading to excess charge density at the apex of the cone geometry. This difference in curvature between spherical and conical geometries leads to variations in charge density in the corresponding systems.
 Fig.~\ref{fig:density}(A) shows the radial density profile for 10-electron $\nu = 1/3$ Laughlin state with $\beta \in \left[1,10 \right]$. $\beta = 1$ corresponds to the plane disc with no curvature and thus the density at the apex equals to the bulk density value $2\pi \rho(r) = \nu $.   When $\beta > 1$,  there is charge accumulation around the cone apex as was observed in the bosonic FQH state~\cite{Ying}.  It is worth noting that the density profile right on the cone tip $2\pi \rho(r) \rightarrow \beta$ with increasing its value.  This suggests that the zeroth orbital is fully occupied and thus a crystalline state is form at the apex. For Laughlin state, it is shown that the condition is $\beta > 4$ as shown in the inserted figure where we plot $\beta - 2\pi \rho(0)$ versus $\beta$.

On the contrary, increasing the value of  $\beta$, which involves reducing the surface area for a fixed radius as depicted in Fig.~\ref{fig:sketch}, results in stretching the cone in the radial direction while maintaining the total area invariant. Consequently, as  $\beta$ increases, the edge moves away from the apex point, making it easier for the system to form a universal quantum Hall edge. The density profiles near the edge for different $\beta$s has a crossover behavior with a rescaled radius $r/\sqrt{\beta}$ as shown in  Fig.~\ref{fig:density}(C). The crosspoint exact  locates at $r/\sqrt{\beta} =\sqrt{2N_{orb}}= \sqrt{2(3N-2)}$ which is the position of the physical edge for N-electron Laughlin state in $(3N-2)$ orbits.  Here we also present the similar data in  Fig.~\ref{fig:density}(B)(D) for the Moore-Read state, another interesting trial state for $\nu = 5/2$ FQH liquid which is supposed to have non-Abelian topological order. Similar to the Laughlin state,  excessive density profile still exists at the cone tip with the exact value $ 2\pi \rho(0) = \beta$ while $\beta > 3$ and the density has crossover at its physical edge $r/\sqrt{\beta} = \sqrt{2(2N-2)}$. Moreover, the density at the cone tip has more pronounced oscillations than that of the Laughlin state which demonstrates different geometric responses for different FQH states.

\begin{figure}[htbp]
\center
\includegraphics[width=8.5cm]{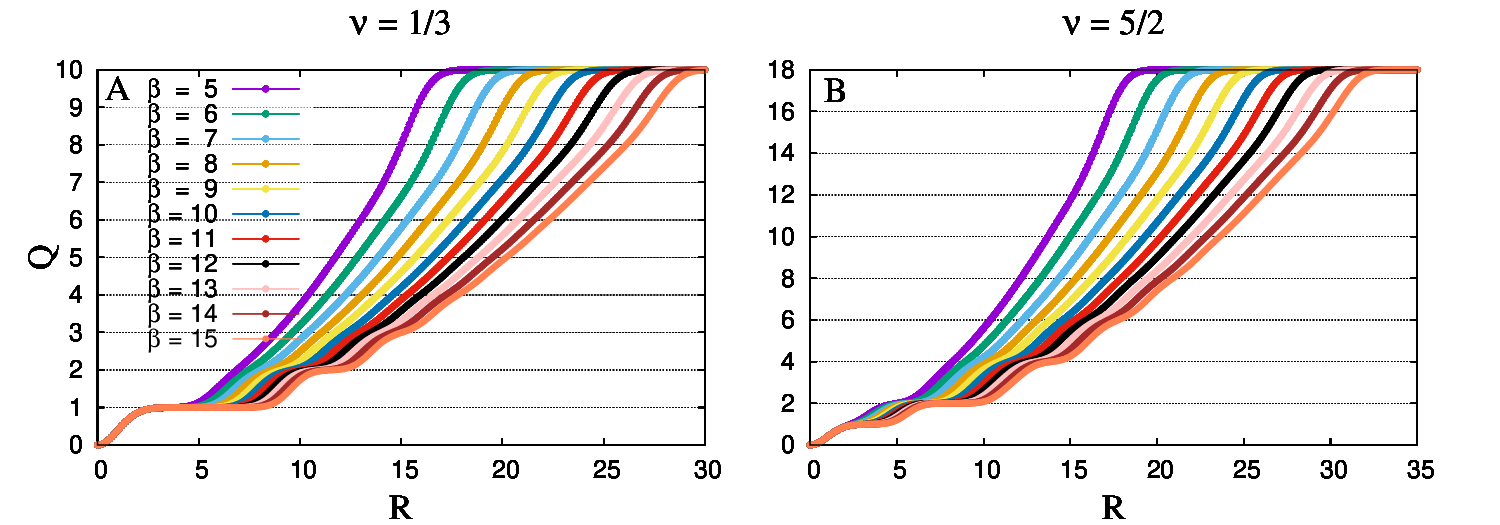}
\caption{The quantity Q in Eq.(\ref{eq:charge}) as a function of the upper limit $R$ of integration for $\nu = 1/3$ Laughlin state (A)  and $\nu = 5/2$ MR state (B) on cones. The system size for Laughlin state is 10 electrons and for MR state  is 18 electrons.}
\label{fig:charge}
\end{figure}

\begin{figure}[htbp]
\includegraphics[width = 8.5cm,height=10cm]{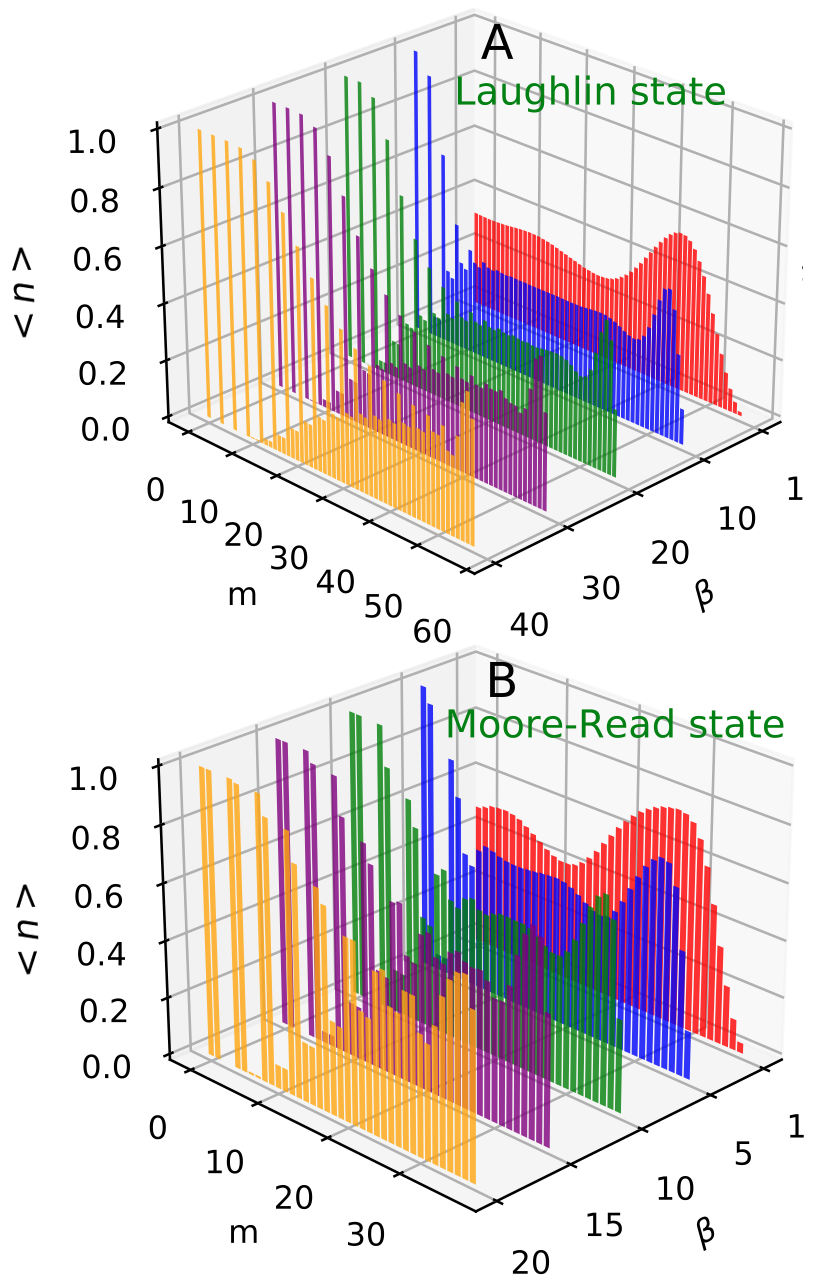}
\caption{The mean orbital occupation number $\langle n \rangle$ for Laughlin state (A)  and MR state (B) with 20 electrons on cones. In each case five discrete $\beta$ values are considered.}
\label{fig:occupation}
\end{figure}

To obtain more detailed information about the charge distribution, we can calculate the accumulated charge over an area that encloses the cone tip in real space. 
\begin{equation}\label{eq:charge}
Q(R)  = \int_0^R \rho(r) \frac{2\pi r}{\beta} dr
\end{equation} 
This calculation will allow us to examine how the charge is distributed within the system and determine if there are any localized regions of excess charge density near the cone apex. As shown in Fig.~\ref{fig:charge}(A) and (B), with increasing the integrated  internal $\left[0,  R \right]$,  the cumulated charge $Q$ starts from zero and increase to the total number of electrons $N$ in the system. As we know,  larger $\beta$ makes the cone thinner and stretch the distance between two nearest electrons analogy to the thin cylinder case. As a result, one could  count the charges more easily with enclosing the integrated area gradually grow. However different from the disk case with smooth ascending curve,  step plateaus emerge start from the cone tip for large $\beta$ cases which indicate the formation of  CDW patterns. Analogy to disk geometry with  symmetric gauge, the Landau orbits on cones are not uniformly distributed thus the charge plateaus in the real space shows different lengths.  Furthermore, we notice the integrated charge step is not always one for Moore-Read state. The first ladder jumps only by one electron charge but the following ladders jump by two electron charges.  Contributions from paired ground state root configuration approximately explains the two steps jumping and the one step jumping could owe to the curvature singularity on the cone tip which always catches one electrons as long as $\beta$ is large enough. However, in the $\beta \rightarrow \infty$ limit, the ladder jumping steps will always be one with two jumps locating closer as a group (``$11$") and the  plateaus  between two  groups  will take longer intervals(``$00$").  The charge pattern could be seen more clearly from the mean orbital electron occupation number $\langle c_n^{\dagger}c_n \rangle=\langle n \rangle$.  The occupation numbers for large system size could be evaluated using Monte Carlo method ~\cite{Morf, Mitra, Khanna} with the help of one-particle reduced density matrix~\cite{Jain}. The technical details for cone geometry are discussed in Appendix~\ref{app:MCocc1/3} and ~\ref{app:MCocc5/2}. As the parameter $\beta$ increases, Fig.~\ref{fig:occupation} illustrates the emergence of an interface that separates the droplets into two distinct regions. In the region close to the cone tip, a charge density wave (TT) phase begins to form, with the leftmost orbital always being occupied. On the other hand, the region near the other end of the cone preserves the fractional quantum Hall (FQH) states.  
 
\textit{orbital angular momentum (OAM)} The OAM of the cone tip should be captured in the net moment  
\begin{equation}
\label{Ltip}
    L_{tip}=\int (\frac{r^2}{2}-1)\Delta \rho(r) dS
\end{equation}
where $\Delta \rho(r)=\rho(r)-\nu/(2\pi)$. The presence of singularity at the origin in the cone geometry leads to the existence of net momentum. This net momentum is a consequence of the gravitational anomaly and is related to important topological quantum numbers. However, if the thermodynamic limit is considered, the net momentum vanishes in the disk geometry. The conformal symmetry has following prediction:\cite{Can,Schine19}
\begin{equation}
	L_{tip}=\frac{c-12\nu \bar s^2}{24}(\beta-\frac{1}{\beta})+\frac{a}{2}(2\bar s-\frac{a}{\beta})
\end{equation}
where $c$ is chiral central charge and $\bar s$ is mean orbital spin of CFT. Here we note that the central charge $c_H$ of Ref. \onlinecite{Can} and the $c$ of Ref. \onlinecite{Schine19} are related by $c_H=c-12\nu \bar s^2$. Therefore, if we consider a Laughlin state at $\nu=1/3, \bar s=\nu^{-1}/2$, the $c=1$ and $c_H=-8$. In the above formula, the first term is brought by the conical tip defect, while the second term is related to quasihole with charge $e/\nu$. Here we consider FQH states without any extra flux threading at the cone tip, in other words, $a=0$. In this case, OAM is reduced to 
\begin{equation}
	L_{tip}=\frac{c-12\nu \bar s^2}{24}(\beta-\frac{1}{\beta})
\end{equation}

\begin{figure}[htbp]
\includegraphics[width=8.5cm]{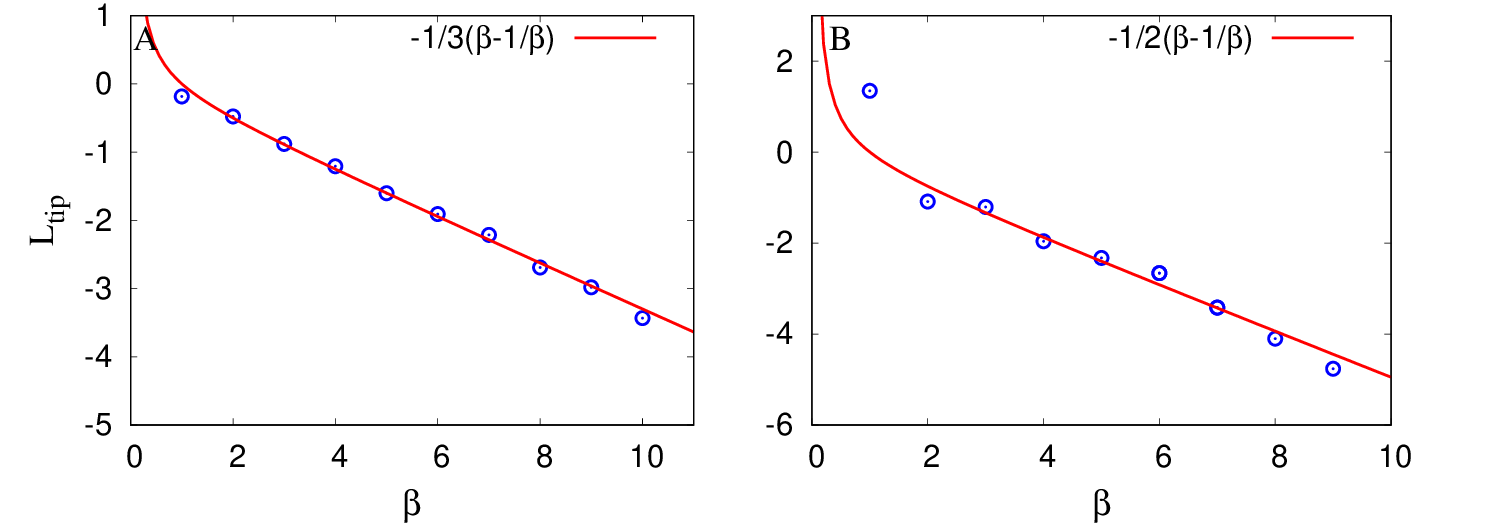}       
\caption{The OAM of the cone tip for 1/3 Laughlin state (A) and MR state (B). The line is theoretical value which is described as $L_{tip}=\frac{c-12\nu \bar s^2}{24}(\beta-\frac{1}{\beta})+\frac{a}{2}(2\bar s-\frac{a}{\beta})$ with $c=1, \bar s=\frac{1}{2}\nu^{-1}, a=0$ for 1/3 Laughlin state and $c=3/2, \bar s=\frac{3}{2}, a=0$ for Moore-Read state.} 
\label{fig:ltip}
\end{figure}

In our numerical calculations, we have determined the $L_{tip}$ for both the Laughlin state and the Moore-Read state through Monte Carlo simulations of a large system with up to 50 electrons. We have utilized an integral upper bound, denoted as $R$, which is positioned far away from both the cone tip and the edge. The results of our calculations are presented in Fig.~\ref{fig:ltip}, where we observe a clear linear relationship between $L_{tip}$  and $\beta-\frac{1}{\beta}$. The fitting slope from our numerical results is in good agreement with the theoretical predictions.

\begin{figure}[htbp]
\includegraphics[width=9cm]{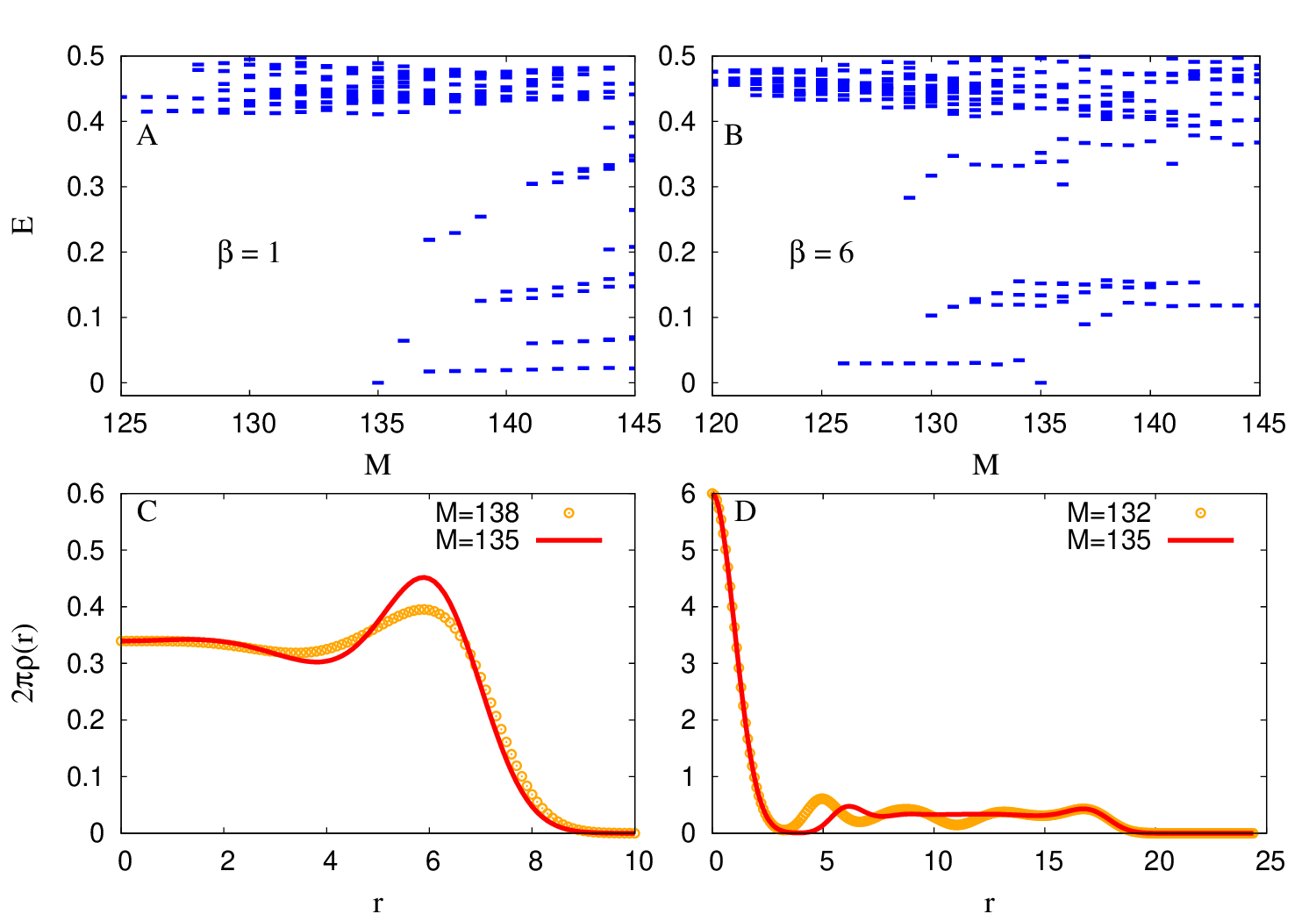}
\caption{The low-lying energy spectrum of a Hard-core interaction with different $\beta=1$(A) and $\beta = 6$(B). The density profile comparison between the Laughlin state and one of the excited states are depicted in (C)and (D). }
\label{spectrum}
\end{figure}

In the following analysis, we will examine the properties of the low-lying excitations in the system. Specifically, we will consider the Laughlin state as an example, which is described by a model Hamiltonian with a hard-core interaction. In this model, only the pseudopotential $V_1$  is non-zero.
For a system with 10 electrons distributed among 28 orbitals, we will investigate the energy spectrum at various values of $\beta$ as shown in Fig.~\ref{spectrum}. The ground state of the system occurs at a total angular momentum $M_{tot} =M_0= 3N(N-1)/2=135$ , which corresponds to zero energy. By examining the energy spectrum at different  $\beta$ values, we can gain insights into the behavior of the system and the nature of its low-lying excitations.  In the case of the plane disk with  $\beta = 1$, the lowest low-lying excited states are the chiral edge excitations, which have a total angular momentum $M_{tot}$ greater than the ground state angular momentum $M_0$. However, as we increase the parameter $\beta$, we observe that these states are gradually raised in energy, and some of the energy levels in the $M_{tot} < M_0$ region are suppressed. This leads to the evolution of these suppressed energy levels into the lowest excited states for $\beta > 5$. Interesting, this is exactly the criteria for developing the interface as discussed previously in Fig.~\ref{fig:density}(A). In Fig.~\ref{spectrum}(B), we can observe that for $\beta = 6$, nearly degenerate energy levels in the range  $M_{tot} \in [126,134]$ are developed as the low-lying excitation branch. This behavior reflects the influence of the parameter $\beta$ on the energy spectrum and the emergence of new low-lying excitations in the system. Fig.~\ref{spectrum}(C)(D) show the comparison of the radial density between the ground state $M_0 = 135$ and one of the lowest  excited state at $M_0 \pm 3$. Obviously, for $\beta = 1$, the $M=138$ is indeed the edge excitation which has a density perturbation  near the edge. Conversely, for $\beta = 6$, the $M=132$ state has a density perturbation in the bulk and keep the density at the cone tip and edge invariant. This could be explained as the interface excitation which has lower energy comparing to the edge excitation. Here we note that as we further increase $\beta$, the excitation energy branch of the interface continues to be suppressed and eventually becomes a zero-energy branch.

\section{wave function profiles}

\begin{figure}[htbp]
\includegraphics[width=8.5cm]{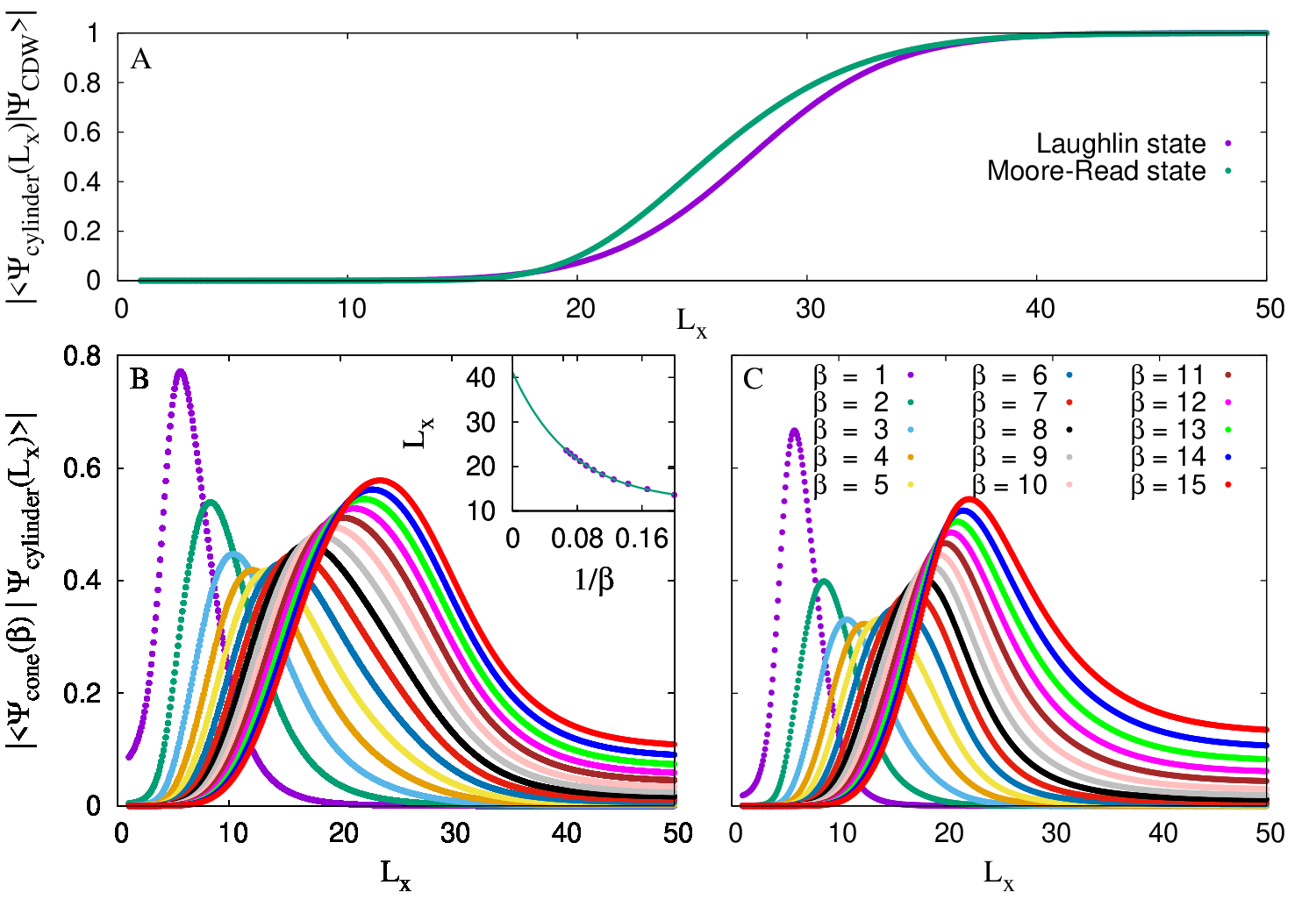}
\caption{(A)The overlaps between wavefunctions on cylinders $\vert \Psi_{cylinder}(L_x) \rangle$ and the CDW state as a function of the length $L_x$. Both Laughlin state and MR states are considered.
The overlaps  $\vert \langle \Psi_{cone}(\beta) \vert \Psi_{cylinder}(L_x) \rangle \vert$ between wavefunctions on cones $\vert \Psi_{cone}(\beta) \rangle$ and  wavefunctions on cylinders $\vert \Psi_{cylinder}(L_x) \rangle$ are plotted in (B) for Laughlin state with 10 electrons and (C) for MR state with 16 electrons. The inset plot in (B) gives the extrapolation of the overlap peak positions $L_x$ for large $\beta$ cases of the Laughlin state.  The corresponding $L_x$ in $\beta \rightarrow \infty$ limit is about $41.1 \ell_B$ for Laughlin state and $34.2 \ell_B$ for MR state.}
\label{fig:overlap}
\end{figure}
Intuitively, for large enough $\beta$, the cone is extremely stretched and  resembles the thin cylinder~\cite{Rezayi1, Seidel, Bergholtz} limit. In order to specify the continue transition to the CDW Tao-Thouless (TT) state~\cite{Tao}, we describe the overlap between wave functions on a cone with varying $\beta$ and that on a cylinder with varying the circumference $L_y$ for the same system size.  In Fig.~\ref{fig:overlap} we plot the corresponding overlaps $\vert \langle \Psi_{cone}(\beta) \vert \Psi_{cylinder}(L_x) \rangle \vert$ for both Laughlin state and MR state. As we know, it only describes an incompressible fluid when two lengths of the cylinder are comparable and when $L_y \rightarrow 0$ or $L_x \rightarrow \infty$ , the ground state is a gapped crystal, the TT state.  In our numerical tests in Fig.~\ref{fig:overlap}(A), the overlap between the ground state on cylinder and the CDW states is already approaching 1 when $L_x \approx 40 \ell_B$ for finite systems. Here in (B) we observe the wavefunction overlaps between cones and cylinders  asymptotically draw near 1 in spite of the finite $\beta$ values and the extrapolation of overlap peaks positions in $\beta \rightarrow \infty $ limit are approaching $41.1 \ell_B (overlap 99.3\%)$ for Laughlin state and $34.2 \ell_B (overlap 92.6\%)$ for MR state.  This results verifies that the $\beta \rightarrow \infty$ limit equals to the thin cylinder limit and the state is indeed the Tao-Thouless state.

 \begin{figure}[htbp]        
\center{\includegraphics[width=8.5cm]  {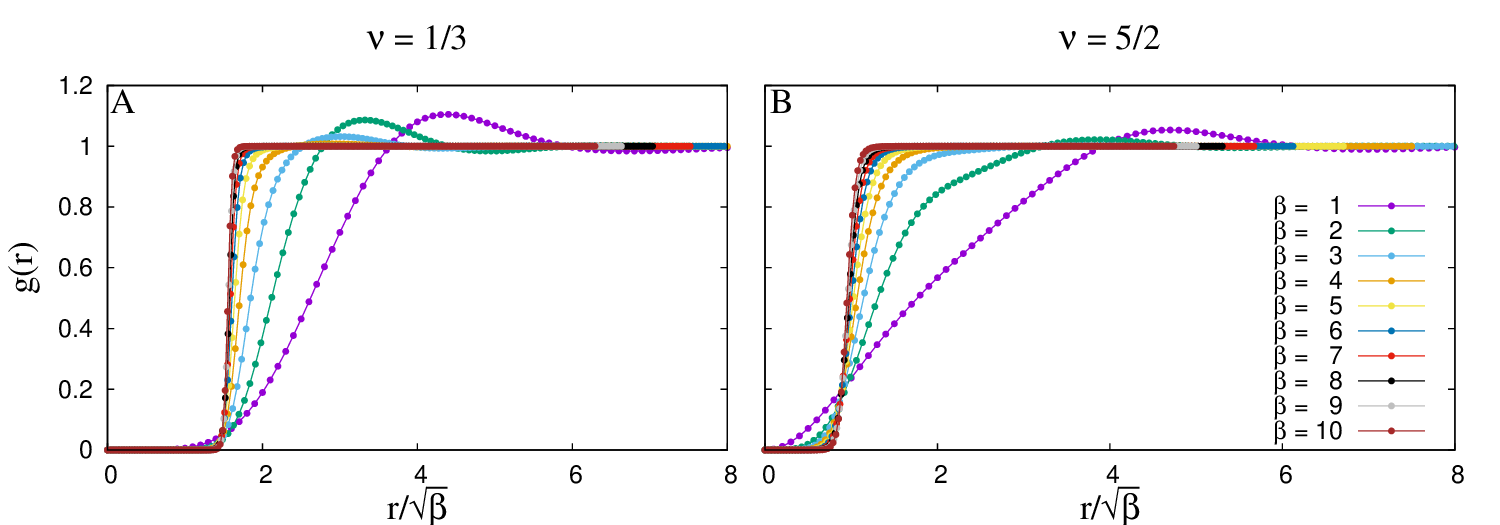}}        
\caption{\label{fig:pair} Pair correlation function $g(r)$ for $1/3$ Laughlin state and $5/2$ MR state for different $\beta$ on a cone. The system size for Laughlin state is 10 electrons and for MR state  is 16 electrons.} 
\end{figure}
\textit{Pair correlation function} In order to investigate the evolution of the electron density near the cone tip, we consider the two-point pair correlation function which is defined as
\begin{eqnarray}
 g(\vec{r}) = \frac{L_xL_y}{N (N - 1)} \langle \Psi | \sum_{i \neq j} \delta(\vec{r} + \vec{r_i} - \vec{r_j})|\Psi\rangle 
\end{eqnarray}
While the coordinate of one particle is fixed at the tip, the pair correlation function can be written in a second quantized form as
\begin{eqnarray} 		
g(r)&=&\frac{1}{\rho(0)\rho(r)}\sum_{m_i} \Phi^*_{\beta m_1}(r) \Phi^*_{\beta m_2} (0) \nonumber \\
 && \Phi_{\beta m_3}(0) \Phi_{\beta m_4} (r) \langle   c^\dagger_{m_1}c^\dagger_{m_2} c_{m_3} c_{m_4}\rangle  
 \end{eqnarray}
It can be obtained using either wave function from diagonalization or Monte Carlo simulation\cite{Orion}.  
 The results are shown in Fig.~\ref{fig:pair} for both the Laughlin state and Moore-Read state in the rescaled radial distance.  In both cases, the $g(r)$ evolves into a sharp step shape.  Taking the Laughlin state as an example, the pair correlation function  $g(r)$ in a plane disk ($\beta = 1$)  exhibits oscillations, characteristic of a liquid-like state. These oscillations are gradually suppressed as the value of  $\beta$ is increased. The peak of $g(r)$, which is larger than 1, disappears at around $\beta = 5$, indicating the formation of a crystalline state near the cone tip, which is consistent with the results from the electron density. Similar analysis applies to the Moore-Read state as shown in Fig.~\ref{fig:pair}(B).

\textit{Entanglement} An effective  tool to extract topological information from the ground state wave function of the FQH states is the entanglement spectrum~\cite{Li} which goes beyond the traditional Landau theory based on  symmetry breaking and local order parameters. To be more precisely, we consider the bipartite entanglement when the Hilbert space is divided into two parts $\mathcal{H} = \mathcal{H}_A + \mathcal{H}_B$.  This partition is characterized by the reduced density matrix $\rho_A = Tr_B \vert  \Psi_0 \rangle \langle \Psi_0 \vert $ after tracing out the degrees of freedom of B.  The bipartite operation on ground state $\Psi_0$ can be implemented in momentum space~\cite{Haque, Zozulya} or  alternatively in real space~\cite{Dubail, Sterdyniak} of the two-dimensional system.  The former is called the orbital cut (OC) and the latter the real space cut (RC). One can perform Schmidt decomposition on $\Psi_0$ and expressed as:
$\vert  \Psi_0 \rangle  = \sum_i e^{-\xi_i/2} \vert \psi_i^A  \rangle  \otimes   \vert \psi_i^B \rangle $
where $ \vert \psi_i^A \rangle $ and $\vert \psi_i^B \rangle $ are orthonormal basis providing a natural bipartition of the system.  The singular values set $e^{-\xi_i/2}$ reveals the  entanglement ``energies" $\xi_i$ which was initially introduced by  Li and Haldane~\cite{Li}. 
As an entanglement measurement, the entanglement entropy is defined associated with $\rho_A$, i.e, the Von Neumann entropy reads $ S_A = - Tr_A \left[ \rho_A \ln \rho_A \right]$.  For two dimensional topological systems,  the entanglement entropy satisfies the area law with a first correction which is named as the topological entanglement entropy $\gamma$~\cite{Hamma, Kitaev, Levin}.  $L$ is the boundary length between two systems in two dimensional case.  $\alpha$ is a non-universal number depends on the way of the bipartition. $S \simeq \alpha L - \gamma$
As a topological order, $\gamma = \ln \mathcal{D}$ is related to the total quantum dimension $\mathcal{D} $ characterizing the topological field theory associated with the phase and the nature of the system excitations.  As we know,  the quantum dimension characterizes the growth rate of the Hilbert space with anyon number and for the fermionic Laughlin state  with anyonic excitations and filling fraction $\nu = 1/m$ it reads $\mathcal{D} =\sqrt{m}$. In addition, when a topological excitation or quasiparticle is emerging in the system, we can detect the quantum dimension of the quasiparticle $d_{\alpha}$ using the additional change of  topological  entanglement entropy $\gamma^{qp} = \ln \mathcal{D} - \ln d_{\alpha} $. In general, the quantum dimension $d_{\alpha} = 1$ for Abelian quasiparticles but $d_{\alpha} > 1$ for non-Abelian quasiparticles.  

\begin{figure}[htbp]
\center
\includegraphics[width=8.5cm]{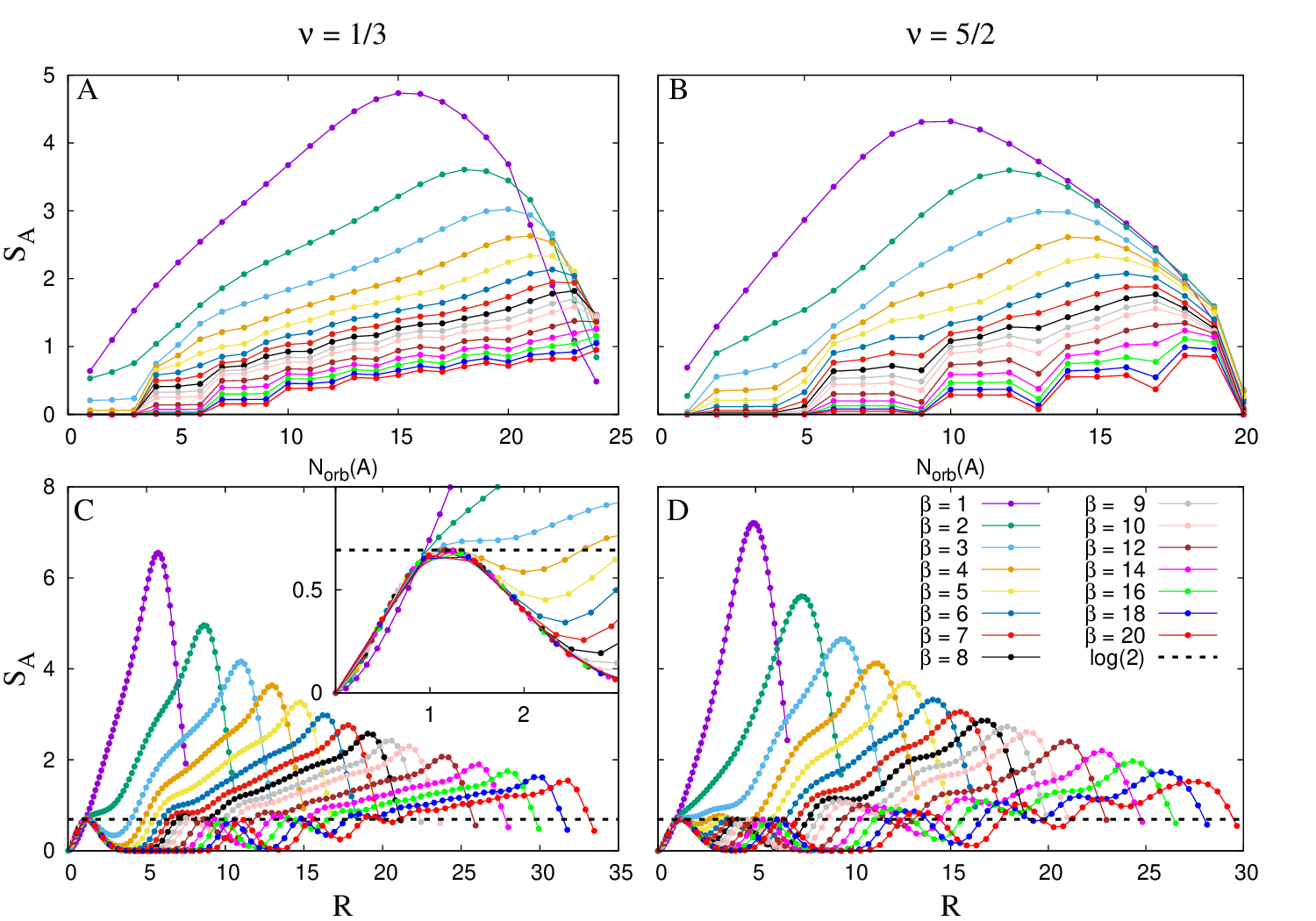}
\caption{The orbital cut entanglement entropy $S_A$ as a function of $N_{orb}(A)$  for Laughlin state (A) and MR state (B) on cones. The total Landau orbital in A subsystem $N_{orb}(A)$ varies as we cut  the system in momentum space in different positions.
The real space cut entanglement entropy $S_A$ as a function of cutting position $R$ for Laughlin state (C) and for MR state (D) on cones. The system size for Laughlin state (A)(C) is 10 electrons and for MR state (B)(D) is 12 electrons. The inset plot in (C) is the enlarged data $S_A$ for  $R \in \left[0, 3 \right]$. The dashed line is  exactly the classical Von Neumann entropy $S = \log(2)$. }
\label{fig:entropy}
\end{figure}

In this work, we study the entanglement entropy of the Laughlin state and MR state on cones for both OC and RC. We focus on the influence of a CDW phase emergence on entanglement entropy in real space and momentum space. For OC it is a natural perspective to vary the cutting position through changing the number of Landau orbital in A subsystem $N_{orb}(A)$. Here we define the left most $N_{orb}(A)$ consecutive orbitals belong to A part which corresponds to the inner circle of a disk or the upper part of a cone. A part still remains the same shape as the whole system but differs in orbital numbers. Here we focus on the cone case especially when large singularity acts on the tip.  With increasing $\beta$, we find the global entropy $S_A$ in Fig.~\ref{fig:entropy}(A)(B) monotonically decrease. When $\beta > 4$,  entropy $S_A$ drops near zero for $N_{orb}(A) = 1,2,3$ in Fig.~\ref{fig:entropy}(A). This means the cone tip lost its correlation with the bulk while the crystalline state is formed. Interestingly,  three consecutive data points comprise a set with almost the same $S_A$ which forms step-like structure,  such as the $\beta = 14$ case forms three steps which is consistent to the occupation pattern $``1001001001\cdots"$ in TT state with three consecutive orbitals as being a unit cell.  Cutting at the left most orbitials with $N_{orb}(A) = 1,2,3$ shares one thing in common, i.e., the A part owes one electron. Once cutting at  $N_{orb}(A) = 4,5,6$, A part owes two electrons  which leads to a new step. For MR state in Fig.~\ref{fig:entropy}(B), large $\beta$ induce CDW phase with configurations $``11001100110011\cdots"$. Transparently, $N_{orb}(A) = 1$ and $N_{orb}(A) = 2$ corresponds to different $N_A$ cases while $N_{orb}(A) = 2,3,4$ form a step with almost equal $S_A$.   Two followed upstairs occur at $N_{orb}(A) = 5$ and $N_{orb}(A) = 6$ with $N_A = 3,4$ respectively.  

For RC case, by cutting the cone  along the loop paralleling to the basal circumference in real space,  a smaller cone defined as part A and the residual part defined as part B are obtained.  The  generatrix  length  $R$ of the smaller cone (part A)  is determined by the real space cutting position and we plot the entanglement entropy $S_A$ against $R$ for Laughlin state and MR state in Fig.~\ref{fig:entropy}(C) and Fig.~\ref{fig:entropy}(D) , respectively. Firstly,  we notice the figures share a common feature that near the cone tip  all entropies  almost collapse into each other (except $\beta = 1$ case).  The enlarged figure inserted in Fig.~\ref{fig:entropy}(C) also shows that for the cone cases with $\beta > 4$, a first peak occurs around $R = 1.18 \ell_B$ with the entropy exactly equals to the classical Von Neumann entropy $S = \log(2)$. This phenomenon implies that the cutting  is right on one electron and all the patterns for big enough $\beta$ cases are almost the same near the cone tip which again shows consistency  for local characteristic length around $1 \ell_B$. In addition,  with increasing the $\beta$ value, a second peak will appear with $S_A > \log(2)$ at larger $R$ and finally equals to $\log(2)$ (cut on electrons again) but never less than $\log(2)$. In the $\beta \rightarrow \infty$ limit, there would have $N-1$ peaks (except the tip one) with values $\log(2)$ which totally corresponds to the CDW phase and is analogous to the thin cylinder case. But within finite $\beta$ case, the CDW pattern begins from the cone tip side and has a evolution process to  fully expand to the whole cone. Compared to the Laughlin state case, the MR  state entropy curves in Fig.~\ref{fig:entropy}(D) show similar behaviors but with first two peaks closer. Here we should note that the first peak occurs also around $R = 1.18 \ell_B$. Thinner cone will extend the distance between two neighbour orbitals and lower the correlations or entropy values between two subsystems. 

\begin{figure}[htbp]
\center
\includegraphics[width=8.5cm]{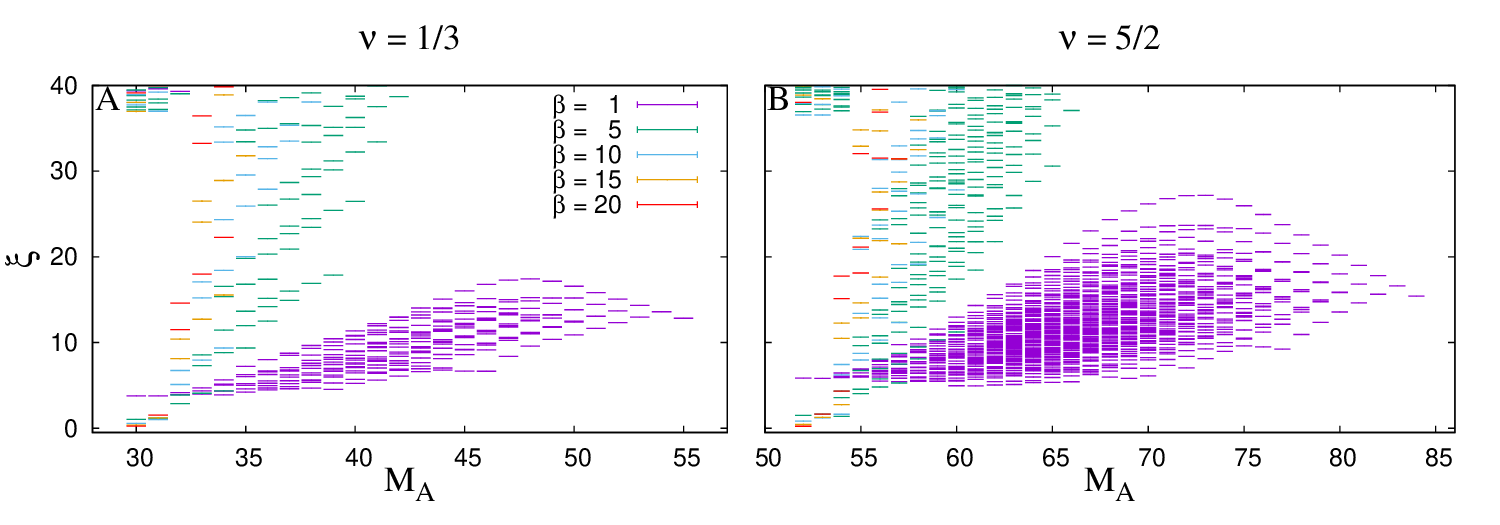}
\caption{The orbital cut entanglement spectrum $\xi$  for Laughlin state (A) and MR state (B) on cones.  The system size for Laughlin state is 10 electrons and for MR state  is 16 electrons.  }
\label{fig:entropyspectrum}
\end{figure}

Fig.~\ref{fig:entropyspectrum}  illustrates the entanglement spectrum of the one-cone  state for half of the system at different values of $\beta$. Notably, the structure of the entanglement spectrum, i.e., the number of states in each momentum space, remains unchanged for all $\beta$ values. This suggests that increasing 
$\beta$ does not lead to a phase transition, indicating that the topological phase of the TT state is the same as that of the Laughlin state. The only variation observed is the steepness of the spectrum. In the TT state, the entanglement is predominantly influenced by the unique ground state in the entanglement spectrum.

\textit{Edge Green's function}
The FQH edge states exhibit a non-Ohmic $I-V$ relation $I \propto V^\eta $ in tunneling experiments, in contrast to the non-interacting Fermi-liquid. This behavior can be predicted by chiral Luttinger liquid theory, and has been observed in experiments~\cite{Chang}. The parameter  $\eta$ in the  $I-V$  relation is a topological quantity of the FQH liquid, with values such as $\eta = 3$ for the $\nu = 1/3$ Laughlin state and the Moore-Read state, as predicted by chiral Luttinger liquid theory~\cite{Wen1,Wen2}. When considering a conical manifold, the edge of the FQH liquid is located far from the tip where the curvature singularity exists, thereby making the edge physics unaffected by the geometric singularity. Moreover, as the system transitions into the TT state by increasing $\beta$, which is topologically equivalent to the Laughlin state, the exponent  $\eta$ should remain constant if it is indeed a topological invariant.

Numerically, the exponent $\eta$ could be obtained from the equal-time edge Green's function $G_{edge}(\vert z_1 -z_2 \vert) \sim \vert z_1 -z_2 \vert^{-\eta}$.  In a system with rotational symmetry, the edge Green's function can be described as 
\begin{eqnarray}
&& G_{edge}(\vert \vec{z_1}- \vec{z_2} \vert) = \Phi^{\dagger }(\vec{z_1}) \Phi (\vec{z_2}) = \nonumber \\
&& \sum_m \frac{\beta }{2\pi 2^{\beta m} \Gamma(\beta m+1) } z_1^{\beta m} z_2^{\beta m} e^{i\beta m(\theta_1-\theta_2)} e^{-\frac{z_1^2+z_2^2}{4}} n_m \nonumber
\end{eqnarray}
two points $\vec{z_1}$ and $\vec{z_2}$ are chosen at the edge of a cone, at a distance of $\vert \vec{z_1} \vert = \vert \vec{z_1} \vert = R=\sqrt{2N_{orb}\beta }+X$ from the tip, where  $\theta=\theta_1-\theta_2$ is the angle between $\vec{z_1}$ and $\vec{z_2}$ and $X$ is the length of density tail.
thus the edge Green's function is rewritten as: 
\begin{equation}
	\Phi^{\dagger }(\vec{z_1}) \Phi (\vec{z_2})=\sum_m \frac{\beta}{2\pi 2^{\beta m} \Gamma(\beta m+1)} R^{2\beta  m} e^{i\beta m\theta}e^{-\frac{R^2}{2}} n_m
\end{equation}

Analytically the chord length between two edge points reads $|\vec{z_1}-\vec{z_2}|=2R/\beta \sin({\theta \beta/2})$ with $\theta \in [0,2\pi/\beta]$.
\begin{figure}[htbp]       
\center
\includegraphics[width=8.5cm]{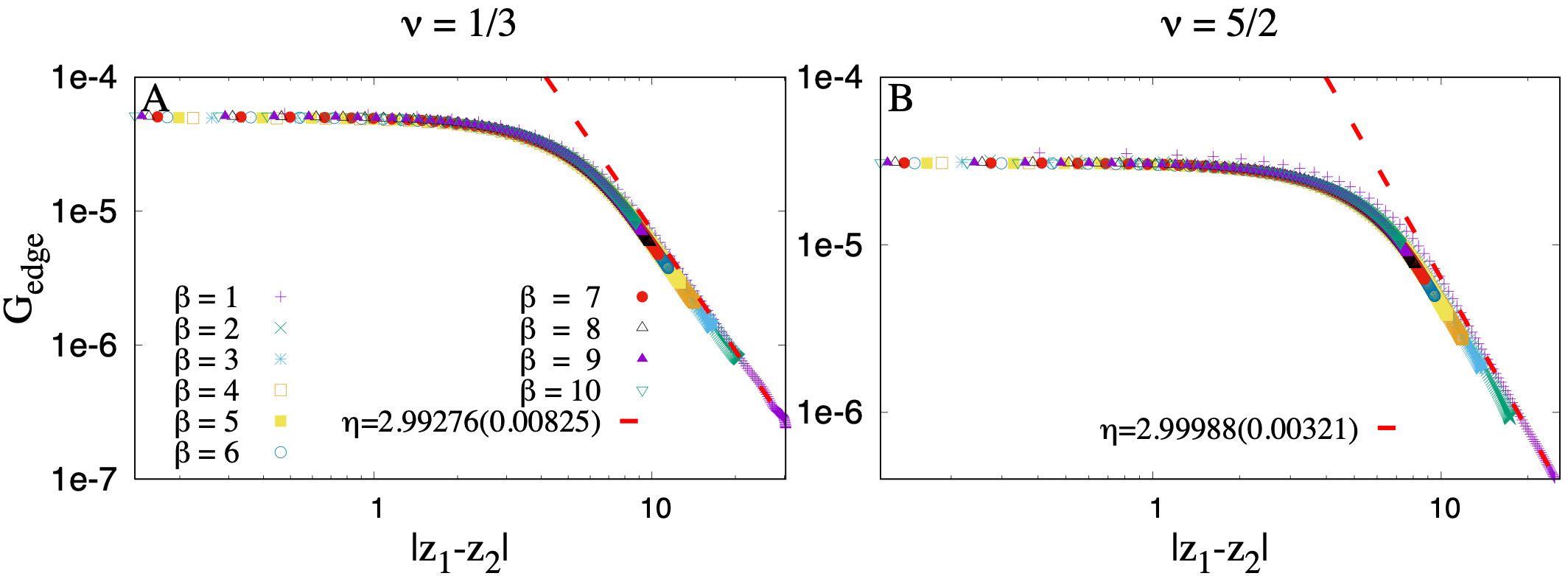}       
\caption{Edge Green's function for $1/3$ Laughlin state and $5/2$ MR state for different $\beta$ with $N=30$ on a cone.  Here we set $X=1.9\ell_B$ and $2.13\ell_B$ respectively for $1/3$ Laughlin state and MR state. The red dashed line is the fitting results. } 
\label{fig:Green}
\end{figure}

As depicted in Fig.~\ref{fig:Green}(A) and (B), or cones with modified curvature, a perfect fitting exponent $\eta \simeq 3$  for both the Laughlin and Moore-Read states while the distance $|z_1 - z_2|$ is large enough, which aligns with the theoretical prediction. Even though a stretched cone has a smaller bottom surface radius, limiting the distance between the two electrons, the edge state of the conical surface still displays the same topological property as the FQH state. The results of the edge green's function still manifest the topological equivalent between the FQH state and its Tao-Thouless limit.

\section{Conclusion}
In summary, our investigation of fractional quantum Hall (FQH) states on conical manifolds has uncovered the formation of a smooth interface that separates the topological trivial (TT) state and the FQH liquid by continuously adjusting the curvature singularity at the tip. The presence of a localized geometrical defect on the cone tip leads to the accumulation of charge due to positive curvature, resulting in a significant modification of the density profile around the apex. The TT state emerges as a signal of a fully occupied zeroth orbital, while for the Laughlin state, $\beta > 4$, and for the Moore-Read state, $\beta > 3$. As the interface between the FQH state and the charge density wave (CDW) state forms, the low-energy spectrum becomes dominated by the density oscillation near the interface, rather than the edge excitation of the FQH liquid. This suggests that interface excitations could play a dominant role in the low-energy physics in realistic scenarios, such as the low-lying excitations of a FQH liquid in sharp confinement and non-uniform electron density. Our orbital angular momentum (OAM) calculations align well with theoretical predictions, demonstrating clearly the gravitational anomaly arising from the geometric singularity. However, through considerations of wave function overlap, entanglement spectrum, and edge Green's function, we confirm that the FQH state and its TT limit are indeed in the same topological phase, indicating that the interface in our work behaves more like a crossover.

\section{Acknowledgments}
The work is supported by National Natural Science Foundation of China Grant No.11974064, 12147102 and No. 61988102; Guangzhou Basic and Applied Basic Research project No. 2023A04J0018. The Key Research and Development Program of Guangdong Province Grant No. 2019B090917007,  the Science and Technology Planning Project of Guangdong Province Grant No. 2019B090909011 and Guangdong Provincial Key Laboratory Grant No. 2019B121203002.  ZL acknowledges the support of funding from Chinese Academy of Science E1Z1D10200 and E2Z2D10200; from ZJ project 2021QN02X159 and from JSPS Grant No. PE14052 and P16027. ZXH also acknowledges Chongqing Talents: Exceptional Young Talents Project No. cstc2021ycjh-bgzxm0147, ChongQing Natural Science Foundation cstc2021jcyj-msxmX0081 and the Fundamental Research Funds for the Central Universities under Grant No. 2020CDJQY-Z003. 
\appendix

\begin{widetext}

\section{Occupation Number of 1/3 Laughlin State from Monte Carlo simulation}
\label{app:MCocc1/3}
In this Appendix, we use Metropolis Monte Carlo (MC) simulation to get the occupation number of FQH states on a cone.
Comparing the single particle wavefunctions Eq.~\eqref{eq:wf} on cone and disk, the (unnormalized) wave-function corresponding to 1/3 Laughlin state $|\Psi_{1/3} \rangle$ is
\begin{equation}
\label{eq3}
	|\Psi_{1/3} \rangle=\prod_{j<k}(\vec{z_j}^{\beta}-\vec{z_k}^{\beta} )^3 e^{-\frac{1}{4}\sum_i z_i^2}
\end{equation}
where $\vec{z}_j=x_j+iy_j=z_je^{i\theta_j}$ is the coordinate of the $j'th$ particle, $\theta_j\in[0,\frac{2\pi}{\beta}] $. 

\par\ The occupation number of $m_{th}$ single-particle orbit of $|\Psi_{1/3} \rangle$ is
\begin{equation}
	\begin{split}
		 n_m^{1/3}&=\frac{\langle \Psi_{1/3} |c_m^{\dagger}c_m|\Psi_{1/3} \rangle  }{\langle \Psi_{1/3}|\Psi_{1/3} \rangle} =\int d \vec{z_1} d \vec{z_2} \rho_{1/3}(\vec{z_1},\vec{z_2})\Phi_{\beta m}^*(\vec{z_1}) \Phi_{\beta m}(\vec{z_2} )
	\end{split}
\end{equation}
where $\rho_{1/3}$ is the one-particle reduced density matrix and $\Phi_{\beta m}$ is the type-I wavefunction of LLL (n=0). $\rho_{1/3}$ can be described as follows\cite{Jain}
\begin{equation}
\label{eq2}
	\rho_{1/3}(\vec{z_a},\vec{z_b})=\frac{N\int \prod_{i=2}^{N} d^2\vec{z_i}\Psi_{1/3}^{*}(\vec{z_a},\vec{z_2},\cdot \cdot \cdot \vec{z_{N}}) \Psi_{1/3}(\vec{z_b},\vec{z_2},\cdot \cdot \cdot \vec{z_{N}} ) }{\int \prod_{i=1}^{N} d^2\vec{z_i}|\Psi_{1/3}|^2 }
\end{equation}
In momentum space the one-particle density matrix can be written as 
\begin{equation}
	\rho_{1/3}(\vec{z_a},\vec{z_b})=\sum_m n_m^{1/3} \Phi_{\beta m}^*(\vec{z_a}) \Phi_{\beta m}(\vec{z_b})
\end{equation}
We choose $\vec{z}_a=ze^{i(\theta_z+\theta)}$ and $\vec{z}_b=ze^{i\theta_z}$ which have the same radial distance but differ by an angle $\theta$ in the complex coordinate system. Thus
\begin{equation}
	\rho_{1/3}(\vec{z_b},\vec{z_b}e^{-i \theta })=\sum_m  n_m^{1/3}  |\Phi_{\beta m}(\vec{z_b})|^2 e^{-i\beta m\theta}
\end{equation}

Then we consider the above relation as a discrete Fourier transform from momentum space index $m$ to real space conjugate $\theta$, and set $\vec{z_b}=\vec{z}$. The inverse transformations read:
\begin{equation}
\label{eq1}
	n_m^{1/3} |\Phi_{\beta m}(\vec{z})|^2 =\frac{1}{3N-2}\sum_{j=0}^{3(N-1)}e^{i\beta m\theta_j}\rho_{1/3}(\vec{z},\vec{z}e^{-i \theta_j })
\end{equation}
where $\theta_j=2\pi j/(3N\beta-2\beta)$. Then we calculate the occupation number by integrating Eq.~(\ref{eq1}) over $\vec{z}$ and get
\begin{equation}
\label{eqocc13}
	n_m^{1/3}=\frac{1}{3N-2}\sum_{j=0}^{3(N-1)}e^{i\beta m\theta_j}\rho_{1/3}(\theta_j)
\end{equation}
where $\rho_{1/3}(\theta_j)=\int d\vec{z} \rho_{1/3} (\vec{z},\vec{z}e^{-i \theta_j})$. Using Eq.~(\ref{eq2}) we have,
\begin{equation}
	\rho_{1/3}(\theta_j)=\frac{N\int \prod_{i=1}^{N} d\vec{z_i}\Psi_{1/3}^{*}(\vec{z_1},\vec{z_2},\cdot \cdot \cdot \vec{z_{N}} ) \Psi_{1/3}(\vec{z_1}e^{-i\theta_j},\vec{z_2},\cdot \cdot \cdot \vec{z_{N}})}{\int \prod_{i=1}^{N} d^2\vec{z_i}|\psi_{1/3}|^2 }
\end{equation}
Using Eq.~(\ref{eq3}) we have,
\begin{equation}
	\Psi_{1/3}(\vec{z_1}e^{-i\theta_j},\vec{z_2},\cdot \cdot \cdot)=\Psi_{1/3}(\vec{z_i})Z_1(\theta_j)
\end{equation}
\begin{equation}
\label{eqza13}
	Z_a(\theta_j)=\prod_{k\neq a}\frac{(\vec{z_a}^{\beta} e^{-i\beta \theta_j}-\vec{z_k}^{\beta})^3}{(\vec{z_a}^{\beta}-\vec{z_k}^{\beta} )^3}
\end{equation}
so we have
\begin{equation}
	\rho_{1/3}(\theta_j)=\frac{N\int \prod_{i=1}^{N} d^2\vec{z_i}|\Psi_{1/3}|^2 Z_1(\theta_j)}{\int \prod_{i=1}^{N} d^2\vec{z_i}|\Psi_{1/3}|^2}=N\langle Z_1(\theta_j) \rangle
\end{equation}
Without loss of generality.
\begin{equation}
\label{eqrho13}
	\rho_{1/3}(\theta_j)=\sum_{a=1}^{N} \langle Z_a(\theta_j) \rangle
\end{equation}
Using Eq.~(\ref{eqocc13}), (\ref{eqza13}) and (\ref{eqrho13}), we finally obtain  the occupation number $n_m^{1/3}$ from MC simulation.

\section{Occupation Number of 5/2 MR State from Monte Carlo simulation}
\label{app:MCocc5/2}
Similarly, the (unnormalized) wave-function corresponding to $\nu = 5/2$ MR state is
\begin{equation}
	\Psi_{5/2}=\textrm{Pf}(\frac{1}{\vec{z}_i^{\beta}-\vec{z}_j^{\beta}} )\prod_{i<j}(\vec{z}_i^{\beta}-\vec{z}_j^{\beta})^2e^{-\frac{1}{4}\sum_i z_i^2}
\end{equation}
where $\textrm{Pf}(Z)$ is the Pfaffian polynomial of matrix $Z$. For instance, in $N=4$ electrons system the matrix $Z$ is equal to
\begin{equation}
 	Z=\begin{bmatrix}
 0 & \frac{1}{\vec{z}_1^\beta-\vec{z}_2^\beta} & \frac{1}{\vec{z}_1^\beta-\vec{z}_3^\beta} & \frac{1}{\vec{z}_1^\beta-\vec{z}_4^\beta}\\
 -\frac{1}{\vec{z}_1^\beta-\vec{z}_2^\beta} & 0 & \frac{1}{\vec{z}_2^\beta-\vec{z}_3^\beta} & \frac{1}{\vec{z}_2^\beta-\vec{z}_4^\beta}\\
 -\frac{1}{\vec{z}_1^\beta-\vec{z}_3^\beta} & -\frac{1}{\vec{z}_2^\beta-\vec{z}_3^\beta} & 0 & \frac{1}{\vec{z}_3^\beta-\vec{z}_4^\beta} \\
 -\frac{1}{\vec{z}_1^\beta-\vec{z}_4^\beta} & -\frac{1}{\vec{z}_2^\beta-\vec{z}_4^\beta} & -\frac{1}{\vec{z}_3^\beta-\vec{z}_4^\beta} & 0
   \end{bmatrix}
\end{equation}
and $\textrm{Pf}(Z)=\frac{1}{\vec{z}_1^\beta-\vec{z}_2^\beta}\frac{1}{\vec{z}_3^\beta-\vec{z}_4^\beta}-\frac{1}{\vec{z}_1^\beta-\vec{z}_3^\beta}\frac{1}{\vec{z}_2^\beta-\vec{z}_4^\beta}+\frac{1}{\vec{z}_1^\beta-\vec{z}_4^\beta}\frac{1}{\vec{z}_2^\beta-\vec{z}_3^\beta}$. While $\textrm{Pf}(Z)$ has a complicated form, its square satisfies $|\textrm{Pf}(Z)|^2=|\det (\textrm{Pf}(Z))|$.
In a similar way, we have
\begin{equation}
	 n_m^{5/2}=\frac{1}{2N-2}\sum_{j=0}^{2N-3}e^{i\beta m\theta_j}\rho_{5/2}(\theta_j)
\end{equation}
\begin{equation}
	\rho_{5/2}(\theta_j)=\sum_{a=1}^{N} \langle Z_a(\theta_j) \rangle
\end{equation}
\begin{equation}
	Z_a(\theta_j)=\prod_{k\neq a}\frac{(\vec{z_a}^\beta e^{-i\beta \theta_j}-\vec{z_k}^\beta )^2}{(\vec{z_a}^\beta -\vec{z_k}^\beta  )^2} \frac{\textrm{Pf}(Z(\vec{z_a}^\beta \rightarrow {(\vec z_a e^{-i \theta_j})^\beta } ))}{\textrm{Pf}(Z)}
\end{equation}
where $\theta_j=2\pi j/ (2\beta N-2\beta)$. We implement the Pfaffian polynomial with the help of the algorithm~\cite{Wimmer}.
\end{widetext}

\section{Accumulated electron Charge for big $\beta$ cases}

\begin{figure}[htbp]        
\center{\includegraphics[width=8.5cm]  {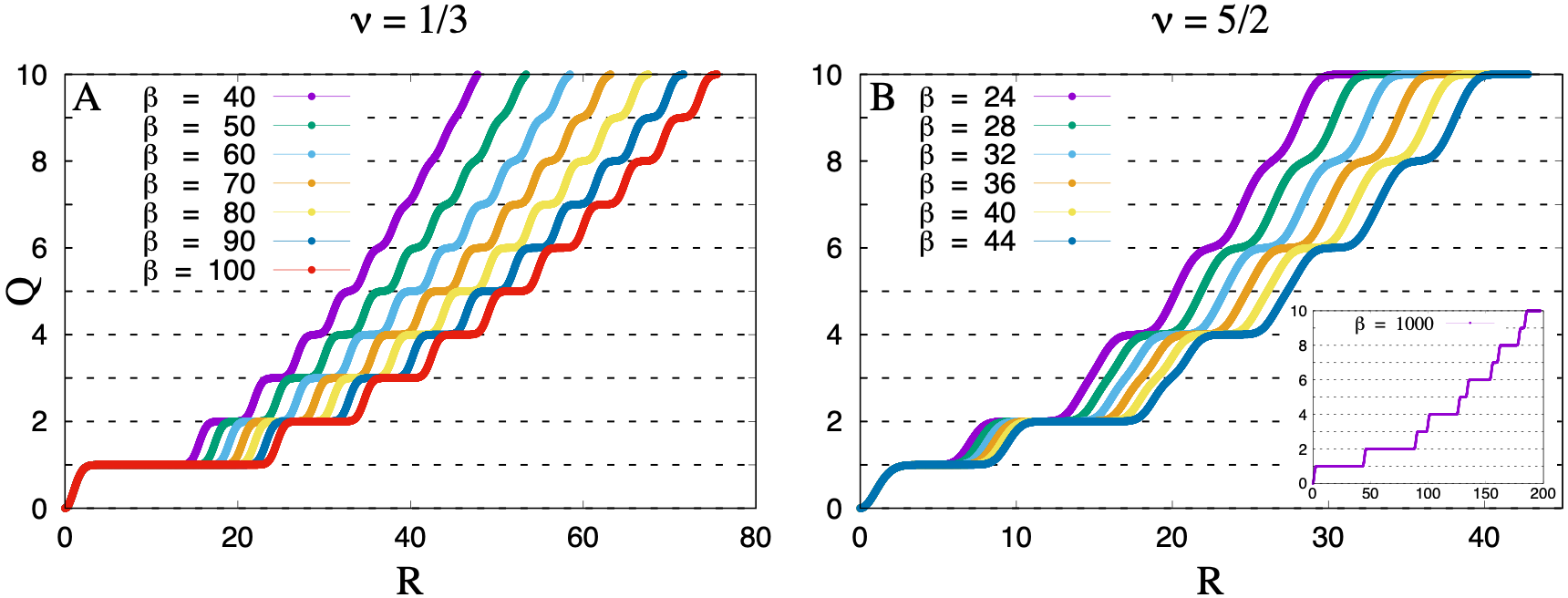}}        
\caption{\label{fig:accbigbeta} Charge accumulation $Q$ from Eq.~\eqref{eq:charge} for $\nu = 1/3$ Laughlin state (A) and  $\nu = 5/2$ MR state (B) with sufficient big $\beta$. The inset in (B) shows the CDW limit for MR state. Every single charge plateau could be seen.} 
\end{figure}

In this Appendix, we supply more numerical results of the accumulated charge for sufficient big $\beta$ cases, as shown in Fig.~\ref{fig:accbigbeta}.

\setcounter{section}{0}
\renewcommand\thesection{\Alph{section}}
\numberwithin{equation}{section}

\end{document}